\documentclass[12pt]{article}

\usepackage{amsmath}
\usepackage{amssymb}
\usepackage{cite}

\def\nn{\nonumber}

\numberwithin{equation}{section}

\title{\bf \Large  New dynamical realizations of the Lifshitz group}

\author{Timofei  Snegirev${}^{a}$\thanks{timofei.v.snegirev@tusur.ru}
\\[0.5cm]
\it{\small ${}^a$Laboratory of Applied Mathematics and Theoretical Physics,}\\
\it{\small Tomsk State University of Control Systems and Radioelectronics,}\\
\it{\small Lenin ave. 40, 634050 Tomsk, Russia}}

\date{}

\begin{document}

\maketitle

\begin{abstract}
The method of nonlinear realizations is applied to construct
new dynamical realizations of the Lifshitz group in mechanics, hydrodynamics, and field theory.
\end{abstract}

\thispagestyle{empty}
\newpage
\setcounter{page}{1}

\section{Introduction}\label{Sec1}

The Lifshitz group \cite{Lif41} (for a review see \cite{MT15})
involves dilatation transformation under which the temporal and
spatial coordinates scale differently: $t'=\lambda^z t$,
$x'_i=~\lambda^{\frac12} x_i$. The parameter $z$ is known as the
dynamical critical exponent. In addition to the scaling
transformations, the group also involves temporal and spatial
translations and spatial rotations. If desirable, Galilei boosts can
also be added. For $z=1$, the group can be also extended by special
conformal transformation, which links it to the Schr\"odinger group.

The Lifshitz group is of interest for several reasons. Originally,
it played an important role in describing phase transitions in
condensed matter physics \cite{Lif41}. More recently, it was used
within the framework of the nonrelativistic AdS/CFT correspondence
\cite{KLM} (for a review see \cite{MT15}). Anisotropic scaling of
temporal and spatial coordinates underlies the Horava-Lifshitz
gravity \cite{Hor08}. Its applications to cosmology are reviewed in
\cite{Muk10}.

When constructing dynamical realizations of the
Lifshitz group in mechanics, a problem arises that the standard free action
$S=\frac12\int dt \dot{x}_i\dot{x}_i$
does not hold invariant under the anisotropic scaling $t'=\lambda^z t$,
$x'_i=~\lambda^{\frac12} x_i$. One way to avoid the difficulty is to raise
the kinetic term to an appropriate power
$(\dot{x}_i\dot{x}_i)^{\frac{z}{2z-1}}$ (see e.g. \cite{RCGV09}). Another option
is to introduce a conformal compensator $\rho(t)$ which transforms with
respect to the dilations as follows $\rho'(t')=\lambda^{z-1} \rho(t)$ \cite{AG22}.
Note that within the conventional group-theoretic framework \cite{CWZ} there is a natural room for
the conformal compensator, where it shows up as a partner accompanying
the dilatation generator in the conformal algebra. In recent works \cite{AG22,FGP22}, the method of
nonlinear realizations  has been successfully applied to build some dynamical realizations of the
Lifshitz group.

The goal of this work is to construct new dynamical realizations of
the Lifshitz group in mechanics, hydrodynamics, and field
theory by applying the conventional group-theoretic framework \cite{CWZ}.

The work is organized as follows. In the next section, we consider a
particular one-dimensional mechanics, for which the spatial
translations and rotations are inessential. In this case, the
scaling symmetry can be realized as a subalgebra of a
one-dimensional conformal algebra $so(2,1)$ \cite{DFF}. It is
demonstrated that the equations of motion have one-parameter
ambiguity and for a special choice they reproduce the recent results
in \cite{AG22}.

In Sect. 3, we consider a dynamical realization of the Lifshitz
group in multidimensional mechanics. Such model describes a particle
moving in $d$-dimensional space driven by a conformal mode, which
can be treated as a dynamical system with varying
mass. It is shown that in a closed system the conformal mode depends only on the
modulus of the conserved momentum, while in a particular isotropic
case $z=\frac12$ it satisfies the equation of motion of the $so(2,1)$ conformal
mechanics in the harmonic trap.

Other many-body mechanics models with the Lifshitz symmetry are discussed in Sect. 4.
Working within
the Hamiltonian framework, conserved charges are constructed which
obey the Lifshitz algebra. Partial differential equations are formulated which
determine an interaction
potential compatible with the Lifshitz symmetry. Their general solution is found. In the
particular case of $z=1$ they reduce to the Calogero-like models studied in
\cite{Cal,AG09}.

In Sect. 5, dynamical realizations of the Lifshitz
algebra, which is extended by the Galilean boost generator as well as
constant accelerations up to some finite order $N$, are discussed. Using the
group-theoretical approach \cite{CWZ}, invariant derivatives and field
combinations are determined, which can be used for model building. It is shown that
in general such models involve higher derivatives.

In Section 6, perfect fluid equations with the Lifshitz symmetry are
studied. Our consideration is similar to the construction of
many-body mechanics. Introducing a conformal mode, one can construct
a Hamiltonian and conserved charges which form the Lifshitz algebra
under the Poisson bracket. In particular, our approach circumvents
the problem of constructing the dilatation generator recently
revealed in \cite{Gal22b}.

In Section 7, it is demonstrated that the
group-theoretical approach can also be used in order to build Lifshitz-type
scalar field theories
\cite{AH07,Vis09,IRS09,EKS11,FIIK15}.

In the concluding Section 8, we summarize our results and discuss possible further developments.

\section{Conformal mode}\label{Sec2}

We start with a simple but instructive example of one-dimensional
mechanics with anisotropic scaling symmetry. Omitting all generators carrying vector indices in
the Lifshitz algebra, one is left with the generator of temporal translation $H$ and dilatation $D$
obeying
\begin{eqnarray}\label{1dLal}
{[H,D]}&=&izH.
\end{eqnarray}
Although in one dimension $z$ can be removed by redefining $D$, keeping in mind forthcoming multi-dimensional generalizations,
we prefer to keep it explicit.

Introducing a temporal variable $t$, the algebra (\ref{1dLal})
can be realized by the operators
\begin{eqnarray*}
H=i\frac{\partial}{\partial t},\quad D=izt\frac{\partial}{\partial
t}.
\end{eqnarray*}
Let us apply the group-theoretic approach \cite{CWZ} to (\ref{1dLal}) with the aim to construct
one-dimensional
mechanics.

Let us consider the group-theoretic element
\begin{eqnarray*}
g=e^{itH}e^{iu(t)D},
\end{eqnarray*}
in which the time variable $t$ is associated with the generator of time
translations $H$ and a field $u(t)$ accompanies $D$. Using the
Baker-Campbell-Hausdorff formula
\begin{eqnarray*}
e^{iA}Te^{-iA}=T+\sum_{n=1}^\infty\frac{i^n}{n!}\underbrace{[A,[A,...[A,T]...]]}_{n~\rm
times}
\end{eqnarray*}
one can calculate the Maurer-Cartan one-forms
\begin{eqnarray}\label{MKform1}
g^{-1}dg=i\omega_{H}H+i\omega_{D}D,
\end{eqnarray}
where
\begin{eqnarray}\label{forms1d}
\omega_H=e^{-uz}dt,\quad \omega_D=du.
\end{eqnarray}
The forms are invariant under the transformation of the group-theoretic element $g'=e^{i\beta
H}e^{i\lambda D}\cdot g$ with real parameters $\beta$ and $\lambda$,
which imply
\begin{align}\label{transf1d}
& t'=t+\beta, && u'(t')=u(t),\nn
\\
& t'=e^{z\lambda}t, && u'(t')=u(t)+\lambda.
\end{align}

Taking into account (\ref{forms1d}), one can build the invariant derivative and field combinations
\begin{eqnarray*}
{\cal D}=\frac{d}{\omega_H}=e^{zu}\frac{d}{dt},\quad {\cal
D}u=\frac{\omega_D}{\omega_H}=e^{zu}\frac{du}{dt}.
\end{eqnarray*}
The latter can be used to construct the equation of motion
\begin{eqnarray}\label{ueq1d}
({\cal D}+a{\cal D}u){\cal D}u=2\gamma^2,
\end{eqnarray}
$\gamma$ and $a$ being arbitrary constants.

Making the field redefinition,
$\rho=~e^{\frac{zu}{2}}$, for which the dilatation transformation reads
$\rho'(t')=e^{\frac{z\lambda}{2}}\rho(t)$, one brings (\ref{ueq1d}) to the form
\begin{eqnarray}\label{1dLeq}
{\ddot\rho}+(\frac{2a}{z}+1)\frac{\dot\rho^2}{\rho}=\frac{z\gamma^2}{\rho^{3}}.
\end{eqnarray}
In order for the corresponding action functional to be invariant
under the dilatation transformation, one has to set
$a=-\frac{z}{2}$ so that the second term in (\ref{1dLeq}) drops
out. The resulting action reads
\begin{eqnarray}\label{1dLac}
S_\rho=\frac12\int
dt\left({\dot\rho}^2-\frac{z\gamma^2}{\rho^{2}}\right).
\end{eqnarray}
Up to a redefinition of the coupling constant, this coincides with the
one-dimensional conformal  mechanics
\cite{DFF}.

By applying the Noether theorem to
(\ref{1dLac}), one gets conserved charges
\begin{eqnarray*}
H=\frac12\left({\dot\rho}^2+\frac{z\gamma^2}{\rho^{2}}\right),\quad
D=\frac{1}{2}\rho\dot\rho-Ht.
\end{eqnarray*}
In order for the energy $H$ to be positive-definite, it is necessary to assume
$z\gamma^2>~0$. Having at our disposal two integrals of motion, one
can construct the general solution to (\ref{1dLac}) by purely
algebraic means
\begin{eqnarray}\label{1dLsol}
\rho^{2}=\frac{4(D+Ht)^2+z\gamma^2}{2H}.
\end{eqnarray}

As compared to the previous studies in \cite{AG22}, where the uniform field redefinition $\rho=~e^{\frac{u}{2}}$
was implemented, in our approach the new field $\rho$ depends on $z$ explicitly which facilitates the explicit
construction of a conserved charge corresponding to the scaling symmetry.
After such a
redefinition, one-dimensional mechanics with the Lifshitz symmetry
coincides with $so(2,1)$ conformal mechanics for arbitrary $z$, not
just for $z=1$ as in \cite{AG22}.

\section{Dynamical realizations of the Lifshitz group in\\  mechanics}\label{Sec3}

Let us now turn to the full Lifshitz
algebra, which in addition to $H$ and $D$ also includes the generators of spatial
translations $C_i$ and spatial rotations $M_{ij}$, $i,j=1,...,d$. The corresponding
structure relations read
\begin{eqnarray}\label{dLal}
{[H,D]}&=&izH,\quad {[D,C_i]}=-\frac{i}{2}C_i,\quad
{[M_{ij},C_k]}=-i\delta_{ki}C_{j}+i\delta_{kj}C_{i},
\end{eqnarray}
$z$ being the critical dynamical exponent. In a non-relativistic space-time
parameterized by $(t,x_i)$, $i=1,...,d$, the algebra can be realized
by the operators
\begin{eqnarray*}
H=i\frac{\partial}{\partial t},\quad D=izt\frac{\partial}{\partial
t}+\frac{i}{2} x_i\frac{\partial}{\partial x_i},\quad
C_i=i\frac{\partial}{\partial x_i},\quad
M_{ij}=ix_i\frac{\partial}{\partial
x_j}-ix_j\frac{\partial}{\partial x_i}.
\end{eqnarray*}

Let us construct multi-dimensional mechanics systems with the
Lifshitz symmetry (\ref{dLal}) by applying the group-theoretic
approach. A similar consideration of the $\ell$-conformal Galilei
group has been given in \cite{FIL11,GM12,AGKM13}

One starts with the coset space element
\begin{eqnarray*}
g=e^{itH}e^{i{x}_i(t)C_i}e^{iu(t)D}\times SO(d),
\end{eqnarray*}
which give rise to the Maurer-Cartan one-forms
\begin{eqnarray*}
g^{-1}dg=i\omega_{H}H+i\omega_iC_i+i\omega_{D}D,
\end{eqnarray*}
where
\begin{eqnarray}\label{formsd}
\omega_H=e^{-uz}dt,\quad \omega_D=du,\quad
\omega_i=e^{-\frac{u}{2}}d{x}_i.
\end{eqnarray}
The forms are invariant under the transformation of the group-theoretic element $g'=e^{i\beta
H}e^{i{a}_iC_i}e^{i\lambda D}\cdot g$ with real parameters $\beta$,
$\lambda$ and $a_i$. The latter implies
\begin{align}\label{Ltr}
& t'=t+\beta, && u'(t')=u(t), && {{x}'}{}_i(t')={x}_i(t),\nn
\\
& t'=e^{z\lambda}t, && u'(t')=u(t)+\lambda, &&
{{x}'}{}_i(t')=e^{\frac{\lambda}{2}}{x}_i(t),\nn
\\
& t'=t, && u'(t')=u(t), && {{x}'}{}_i(t')={x}_i(t)+a_i.
\end{align}
The transformation laws of $x_i(t)$ indicate that they
should be interpreted as describing particle's orbit. From
(\ref{formsd}) one obtains the invariants
\begin{eqnarray}\label{invder}
{\cal D}=\frac{d}{\omega_H}=e^{zu}\frac{d}{dt},\quad {\cal
D}u=\frac{\omega_D}{\omega_H}=e^{zu}\frac{du}{dt},\quad
\frac{\omega_i}{\omega_H}=e^{u(z-\frac12)}\frac{d{x}_i}{dt},
\end{eqnarray}
which can be used to build equations of motion.

For the conformal mode
$\rho=e^{\frac{zu}{2}}$ it seems natural to choose the equation (\ref{1dLac})
 constructed in the previous section. For field
variables $x_i(t)$ we take the most general linear expression
quadratic in derivatives
\begin{eqnarray}\label{MLIeq}
({\cal D}+a{\cal D}u)\frac{\omega_i}{\omega_H}=0,
\end{eqnarray}
where $a$ is a free parameter which will be fixed later. Taking into
account $\rho=e^{\frac{zu}{2}}$ and (\ref{invder}) the equation
(\ref{MLIeq}) can be rewritten as
\begin{eqnarray}\label{xeqd}
\ddot{x}_i+\frac{(2a+2z-1)}{z}\frac{\dot\rho}{\rho}\dot{x}_i=0.
\end{eqnarray}
Similar model was considered in
\cite{AG22}, which describes an oscillator with a time-dependent
frequency and damping in the co-moving frame
$\tilde{x}_i(t)=~\rho^{-\frac {1}{z}}x_i(t)$. Note that $\tilde{x}_i(t)$
remains inert under the dilatation\footnote{The results in \cite{AG22}
are reproduced from the formula (\ref{xeqd}) by going to the
co-moving frame $\tilde{x}_i(t)=~\rho^{-\frac{1}{z}}x_i(t)$ and
redefining $\rho\rightarrow\rho^{z}$.}. In the conventional
coordinate frame $x_i(t)$, the equation (\ref{xeqd}) can be obtained
from the action functional
\begin{eqnarray}\label{MLac}
S_x=\frac12\int dt\rho^{\alpha}\dot{x}_i\dot{x}_i,
\end{eqnarray}
where we redefined the free parameter $\alpha=\frac{(2a+2z-1)}{z}$.
Finally, the parameter $\alpha$ can be fixed by demanding the
dilatation invariance of (\ref{MLac}): $\alpha=\frac{2(z-1)}{z}$.
Note that $R(t)=\rho^\alpha$ can be interpreted as a cosmic scale
factor entering the spatial metric $g^{ij}=R(t)\delta^{ij}$ (see
e.g. the discussion in \cite{AG22}). Alternatively,
$m(t)=\rho^\alpha$ can be regarded as a varying mass.

To summarize, the equations of motion of the multi-dimensional mechanics invariant under the
Lifshitz group read
\begin{eqnarray}\label{MLeq2}
{\ddot\rho}=\frac{z\gamma^2}{\rho^{3}},\qquad
\ddot{x}_i+\frac{2(z-1)}{z}\frac{\dot\rho}{\rho}\dot{x}_i=0.
\end{eqnarray}
The second equation describes the dynamics of a
particle in $d$ spatial dimensions $x_i(t)$, which is driven by the conformal mode
$\rho(t)$ satisfying the first equation. The general solution to the first
equation was found in the preceding section (\ref{1dLsol}), which yields
\begin{eqnarray*}
{x}_i=C_i\int\left(\frac{2H_\rho}{4(D_\rho+H_\rho
t)^2+z\gamma^2}\right)^{\frac{z-1}{z}}dt,
\end{eqnarray*}
where $C_i$ is a constant of integration.

The system of equations (\ref{MLeq2}) is not Lagrangian. In order to obtain a Lagrangian mechanics,
we consider the modified action functional
\begin{eqnarray}\label{req1d}
S=\frac12\int
dt\left({\dot\rho}^2+\rho^{\frac{2(z-1)}{z}}\dot{x}_i\dot{x}_i-\frac{z\gamma^2}{\rho^{2}}\right),
\end{eqnarray}
which gives the equations of motion
\begin{eqnarray}\label{MLeq3}
\ddot\rho-\frac{z-1}{z}\rho^{\frac{z-2}{z}}\dot{x}_i\dot{x}_i=\frac{z\gamma^2}{\rho^{3}},\qquad
\ddot{x}_i+\frac{2(z-1)}{z}\frac{\dot\rho}{\rho}\dot{x}_i=0.
\end{eqnarray}
Noether's integrals corresponding to the Lifshitz
symmetry read
\begin{align}
&
H=\frac12\left({\dot\rho}^2+\rho^{-\alpha}C_iC_i+\frac{z\gamma^2}{\rho^{2}}\right),
&& C_i=\rho^\alpha\dot{x}_i,
\\
& D=\frac12\rho\dot\rho+\frac{1}{2z}\rho^\alpha x_i\dot{x}_i-Ht, &&
M_{ij}=x_iC_j-x_jC_i.
\end{align}
Note that the first equation in (\ref{MLeq3}) does not explicitly
depend on $x_i$, it depends on the square of the conserved momentum
of the system $C_i=\rho^{\frac{2(z-1)}{z}}\dot{x}_i$ and can be
written in the form
\begin{eqnarray}\label{MLeqr}
\ddot\rho-\frac{z-1}{z}C_iC_i\rho^{\frac{-3z+2}{z}}=\frac{z\gamma^2}{\rho^{3}}.
\end{eqnarray}
For $z=1$ the second term on the left-hand side drops out and the
equation describes the $so(2,1)$ conformal mechanics discussed in
\ref{Sec2}. At the same time, for $z=\frac12$ the equation
(\ref{MLeqr}) describes conformal mechanics in a harmonic trap
\begin{eqnarray*}
\ddot\rho+\omega^2\rho=\frac{z\gamma^2}{\rho^{3}},\quad
\omega^2=C_iC_i
\end{eqnarray*}
which is $so(2,1)$-invariant as well. Its general solution describes
oscillations around the equilibrium point
$\rho_0=\sqrt{\frac{\hat\gamma}{\omega}}$ (here
$\hat\gamma^2=z\gamma^2$)
\begin{eqnarray}
\rho^2=\frac{1}{\omega^2}(H-\sqrt{H^2-\omega^2\hat\gamma^2}\cos2\omega(t+t_0))
\end{eqnarray}
and reduces to (\ref{1dLsol}) in the $\omega\rightarrow0$ limit by
defining integration constant $t_0=\frac{D}{H}$. In the case of an
arbitrary $z$ the equilibrium point of the equation (\ref{MLeqr}) is
equal to
$\rho_0=\left({\frac{z\hat\gamma^2}{(1-z)\omega^2}}\right)^{\frac{z}{2}}$
and the general solution can be written in quadratures
\begin{eqnarray*}
\pm(t+t_0)=\int\frac{\rho
d\rho}{\sqrt{2H\rho^2-\omega^2\rho^{\frac{2}{z}}-\hat\gamma^2}}.
\end{eqnarray*}

As compared to \cite{AG22}, we have constructed the complete
Lagrangian dynamical realization of the Lifshitz group (\ref{req1d})
by modifying the equation of motion for the conformal mode
$\rho(t)$, in which $\rho$ is not a background field in
which a particle parametrized by $x_i(t)$ moves, but rather it makes part of a closed system.

In the next section we consider a generalization of the action
(\ref{req1d}) to the case of many-body dynamical systems.

\section{Many-body mechanics with Lifshitz symmetry}\label{Sec4}

In order to construct models of many-body mechanics with the Lifshitz symmetry, it
proves convenient to pass to the Hamiltonian formalism. The phase
space of $n$ identical particles is defined by the canonical pairs
$(\rho,\pi)$ and $(x_i^A,p_i^A)$, $A=1,...,n$ with the standard Poisson
bracket
\begin{eqnarray*}
\{\rho,\pi\}=1,\qquad \{x_i^A,p_j^B\}=\delta_{ij}\delta^{AB}.
\end{eqnarray*}
The dynamics is governed by the Hamiltonian
\begin{eqnarray}\label{MBH}
H&=&\frac12\left(\pi^2+\rho^{-\alpha}p_i^Ap_i^A+\frac{z\gamma^2}{\rho^{2}}+V(\rho,x_i^A)\right),
\end{eqnarray}
where $V(\rho,x_i^A)$ is a potential. Conserved charges corresponding to
dilatation, spatial translations and spatial rotations have the form
\begin{eqnarray*}
D&=&-\frac12z\rho\pi-\frac{1}{2} x_i^Ap_i^A+zHt,\quad
C_i=\sum_Ap_i^A,\quad M_{ij}=x_i^Ap_j^A-x_j^Ap_i^A.
\end{eqnarray*}
Together with the conserved energy $H$ defined by the Hamiltonian
(\ref{MBH}) they satisfy the Lifshitz algebra (\ref{dLal}) under the Poisson brackets
\begin{eqnarray*}
\{H,D\}=zH,\quad \{D,C_i\}=-\frac12C_i,\quad
\{M_{ij},C_k\}=-\delta_{ki}C_{j}+\delta_{kj}C_{i},
\end{eqnarray*}
provided the potential $V(\rho,x_i^A)$ obeys the
first order partial differential equations
\begin{eqnarray}\label{PotEq}
\sum_A\frac{\partial V}{\partial x_i^A}=0,\quad x_i^A\frac{\partial
V}{\partial x_j^A}-x_j^A\frac{\partial V}{\partial x_i^A}=0,\quad
\rho\frac{\partial V}{\partial\rho}+\frac{1}{z}x_i^A\frac{\partial
V}{\partial x_i^A}+2V=0.
\end{eqnarray}
The first two conditions arise from the invariance under spatial
translations and rotations, while the latter guarantees the
dilatation symmetry.

The first equation in (\ref{PotEq}) implies that the potential $V$
must depend on the combinations $x_i^{AB}=x_i^A-x_i^B$. In total,
for an $n$-particle system there are $n-1$ independent combinations
which can be parameterized as $x^{\hat{A}}=x^{1\hat{A}}$,
$\hat{A}=2,3,...,n$. The third equation in (\ref{PotEq}) restricts
the form of the potential as follows
\begin{eqnarray*}
V=\frac{1}{\rho^{2}}F\left(\rho^{-\frac{1}{z}}x_i^{\hat{A}}\right).
\end{eqnarray*}
Finally, in order to take into account the second equation in
(\ref{PotEq}), one has to form $SO(d)$ scalars from
$\rho^{-\frac{1}{z}}x_i^{\hat{A}}$
\begin{eqnarray*}
\upsilon^{\hat{A}\hat{B}}=\rho^{-\frac{2}{z}}\delta^{ij}x_i^{\hat{A}}x_j^{\hat{B}},\quad
\omega^{\hat{A}_1...\hat{A}_d}=\rho^{-\frac{d}{z}}\varepsilon^{i_1i_2...i_d}x_{i_1}^{\hat{A}_1}x_{i_2}^{\hat{A}_2}...x_{i_d}^{\hat{A}_d}.
\end{eqnarray*}
Here $\delta^{ij}$ and $\varepsilon_{i_1i_2...i_d}$ are the
Euclidean metric and the Levi-Cevita symbol in $R^d$ which are
$SO(d)$-invariant. Note that $\omega^{\hat{A}_1...\hat{A}_d}$ is actually a
pseudoscalar and in $d$ dimensions there is a restriction on a
minimal number of particles
$n_{min}=d+1$. Thus, a solution to
(\ref{PotEq}) can be written in the form
\begin{eqnarray*}
V=\frac{1}{\rho^{2}}F\left(\upsilon^{\hat{A}\hat{B}},\omega^{\hat{A}_1...\hat{A}_d}\right).
\end{eqnarray*}

In order to obtain reasonable models from an infinite number of solutions,
one can impose additional restrictions on $V$ such as homogeneity,
parity ($P$-symmetry), permutation symmetry, supersymmetry,
integrability, and etc. Let us give two examples of potentials
which are invariant under permutations
\begin{eqnarray*}
V_1=g_1\sum_{A<B}(x_i^{{A}B}x_i^{A{B}})^{-z},\quad
V_2=g_2\sum_{A<B}\rho^{-\frac{2(z-1)}{z}}}({x_i^{{A}B}x_i^{A{B}})^{-1},\quad
x_i^{AB}=x_i^A-x_i^B,
\end{eqnarray*}
where the first example represents the only possibility of a
potential built independent on $\rho$.
Both examples for $z=1$ reproduce the Calogero-like models in
\cite{AG09}.

\section{Dynamical realizations of the extended Lifshitz group}\label{Sec5}

The Lifshitz algebra can be extended by adding Galilei boost and a
finite number of constant acceleration generators $C_i{}^{(n)}$,
$n=0,1,...,N$. The corresponding structure relations read
\begin{eqnarray}\label{EzLal}
{[H,D]}&=&izH,\quad {[H,C^{(n)}_i]}=inC^{(n-1)}_i,\quad
{[D,C^{(n)}_i]}=i(zn-\frac12)C^{(n)}_i\nn
\\
&&{[M_{ij},C^{(n)}_k]}=-i\delta_{ki}C^{(n)}_{j}+i\delta_{kj}C^{(n)}_{i},
\end{eqnarray}
where $C_i{}^{(n)}$ for $n=0,1$ correspond to spatial translations
and Galilei boosts, while for $n>1$ they link to constant accelerations.
The algebra can be realized by the operators
\begin{eqnarray*}
H=i\frac{\partial}{\partial t},\quad D=izt\frac{\partial}{\partial
t}+\frac{i}{2} x_i\frac{\partial}{\partial x_i},\quad
C^{(n)}_i=it^n\frac{\partial}{\partial x_i}.
\end{eqnarray*}

Repeating the steps in the previous section, one constructs the coset
space element
\begin{eqnarray*}
g=e^{itH}e^{i{x}^{(n)}_i(t)C_i^{(n)}}e^{iu(t)D}\times SO(d),
\end{eqnarray*}
where the summation over $n$ is understood, and derives the
Maurer-Cartan one-forms
\begin{eqnarray}\label{MKform1E}
g^{-1}dg=i\omega_{H}H+i\omega_i^{(n)}C_i^{(n)}+i\omega_{D}D,
\end{eqnarray}
involving
\begin{eqnarray}\label{formacc}
\omega_H=e^{-uz}dt,\quad \omega_D=du,\quad
\omega_i^{(n)}=e^{u(zn-\frac12)}\left(d{x}_i^{(n)}-(n+1){x}_i^{(n+1)}dt\right).
\end{eqnarray}
The forms are invariant under the transformation $g'=e^{i\beta
H}e^{i{a}_i^{(n)}C_i^{(n)}}e^{i\lambda D}\cdot g$ with real
parameters $\beta$, $\lambda$ and $a_i^{(n)}$, which implies
\begin{align*}
& t'=t+\beta, && u'(t')=u(t), && {{x}'}{}_i^{(n)}={x}_i^{(n)},
\\
& t'=e^{z\lambda}t, && u'(t')=u(t)+\lambda, &&
{{x}'}{}_i^{(n)}=e^{-\lambda(zn-\frac12)}{x}_i^{(n)},
\\
& t'=t, && u'(t')=u(t), &&
{{x}'}{}_i^{(n)}={x}_i^{(n)}+\frac{m!}{(m-n)!n!}t^{m-n}a_i^{(m)},\quad
n\leq m,
\end{align*}
where $m=0,1,...,N$.

From  (\ref{formacc}), one gets the invariant derivative and
field combinations
\begin{eqnarray}\label{invderacc}
{\cal D}=\frac{d}{\omega_H}=e^{zu}\frac{d}{dt},\quad {\cal
D}u=\frac{\omega_D}{\omega_H}=e^{zu}\frac{du}{dt},\quad
\frac{\omega_i^{(N)}}{\omega_H}=e^{u(z[N+1]-\frac12)}\frac{d{x}_i^{(N)}}{dt}.
\end{eqnarray}
Imposing the algebraic invariant constraints\footnote{They provide a
simple example of the inverse Higgs constraints \cite{IO75}}
\begin{eqnarray}\label{constr}
\frac{\omega_i^{(n)}}{\omega_H}=0\quad\Rightarrow\quad
(n+1){x}_i^{(n+1)}=\frac{d{x}_i^{(n)}}{dt},\quad n=0,1,...,N-1,
\end{eqnarray}
one readily gets
$$
{x}_i^{(N)}=\frac{d^N}{dt^N}x_i^{(0)}.
$$

By analogy with the case (\ref{MLIeq}), let us consider the equation of
motion
\begin{eqnarray*}
({\cal D}+a{\cal D}u)\frac{\omega_i^{(N)}}{\omega_H}=0,
\end{eqnarray*}
where $a$ is a free parameter. Taking into account
$\rho=e^{\frac{zu}{2}}$, (\ref{invderacc}) and constraints
(\ref{constr}) the equation can be rewritten in the form
\begin{eqnarray}\label{xeqacc}
\frac{d^{N+2}}{dt^{N+2}}x_i^{(0)}+\alpha\frac{\dot\rho}{\rho}\frac{d^{N+1}}{dt^{N+1}}x_i^{(0)}=0,
\end{eqnarray}
where the free parameter was redefined as follows
$\alpha=\frac{2(a+z[N+1]-\frac12)}{z}$. To fix $\alpha$ we need a
suitable action functional. In the particular case $\alpha=0$ the
equation (\ref{xeqacc}) describes a free particle with a constant
acceleration of order $N$, which has a larger $\ell$-conformal
Galilei symmetry for an arbitrary $\ell=\frac{N +1}{2}$
\cite{NOR97}.

In the general case, the equation has the Lifshitz
symmetry extended by the Galilei boosts and constant accelerations.
Demanding $\rho(t)$ to obey (\ref{1dLsol}), the
general solution to (\ref{xeqacc}) can be written in the form
\begin{eqnarray*}
{x}_i^{(0)}=C_i\underbrace{\int ...\int}_{N+1~\rm
times}\left(\frac{2H_\rho}{4(D_\rho+H_\rho
t)^2+z\gamma^2}\right)^{\frac{\alpha}{2}}dt^{N+1},
\end{eqnarray*}
where $C_i$ is the integral of motion
\begin{eqnarray*}
C_i=\rho^{\alpha}\frac{d^{N+1}}{dt^{N+1}}x_i^{(0)}.
\end{eqnarray*}

One more dynamical realization of the extended Lifshitz
algebra (\ref{EzLal}) is described by the invariant action functional
\begin{eqnarray}\label{ExMLac}
S=\frac12\int
dt\rho^{\frac{2(z[2N+1]-1)}{z}}\frac{d^{N+1}}{dt^{N+1}}{x}_i^{(0)}\frac{d^{N+1}}{dt^{N+1}}{x}_i^{(0)}.
\end{eqnarray}
The latter gives rise to the equations of motion
\begin{eqnarray*}
\frac{d^{N+1}}{dt^{N+1}}\left(\rho^{\frac{2(z[2N+1]-1)}{z}}\frac{d^{N+1}}{dt^{N+1}}{x}_i^{(0)}\right)=0
\end{eqnarray*}
which can be rewritten in terms of the invariant objects (\ref{invderacc})
and the constraints (\ref{constr})
\begin{eqnarray*}
\prod_{k=0}^N\left({\cal D}+(z[N-k]-\frac12){\cal
D}u\right)\frac{\omega_i^{(N)}}{\omega_H}=0.
\end{eqnarray*}
Note that (\ref{xeqacc}) and  (\ref{ExMLac}) represent higher derivative
generalizations of (\ref{MLac}) of the order $N+2$ and $2(N+1)$, respectively.

\section{Perfect fluid equations with the Lifshitz symmetry}\label{Sec6}

It is known that perfect fluid equations supplemented with an
appropriate equation of state enjoy the Schr\"odinger symmetry
\cite{JNPP,HZ}. Generalized perfect fluid equations with the
$\ell$-conformal Galilei symmetry and their Hamiltonian formulation
were recently considered in \cite{Gal22a,Gal22b,Sne23a}. In this
section, we discuss perfect fluid equations with the Lifshitz
symmetry.

In a recent work \cite{Gal22b}, the invariant equations were proposed
\begin{eqnarray}\label{FLeq}
\frac{\partial\varrho}{\partial t}+ \frac{\partial
(\varrho\upsilon_i)}{\partial x_i}=0,\quad {\cal
D}\upsilon_i=-\frac{1}{\varrho}\frac{\partial p}{\partial x_i},\quad
p=\nu \varrho^{1+\frac{2(2z-1)}{ d}},
\end{eqnarray}
where ${\cal D}=\frac{\partial}{\partial
t}+\upsilon_i\frac{\partial}{\partial x_i}$ is the material derivative.
The first equation is the continuity equation for fluid density
$\varrho(t,x)$, while the second equation is the Euler equation. The third equation is the equation of state.

Eqs.
(\ref{FLeq}) hold invariant under transformations forming the Lifshitz group
\begin{align}\label{SLT}
& t'=t+\beta,  && {{x}'}{}_i={x}_i, &&
\\
& t'=e^{z\lambda}t, && {{x}'}{}_i=e^{\frac{\lambda}{2}}{x}_i, &&
\\
& t'=t,  && {{x}'}{}_i={x}_i+a_i. &&
\end{align}
where we omitted the conventional spatial rotations.
The fields $\varrho(t,x)$ and $\upsilon_i(t,x)$ transform nontrivially
only under the dilatations (for more details see
\cite{Gal22b})
\begin{eqnarray}
\varrho'(t',x')=e^{-\frac{d}{2}\lambda}\varrho(t,x),\quad\upsilon_i(t,x)=e^{(z-\frac12)\lambda}\upsilon'_i(t',x').
\end{eqnarray}
However, the construction of conserved charge corresponding to dilatation faces a problem.
Below, we construct a one-parameter generalization of (\ref{FLeq}), for which all conserved charges can be
constructed explicitly.

In addition to the Lifshitz symmetry, the equations (\ref{FLeq})
hold invariant under the Galilei boosts ${{x}'}{}_i={x}_i+b_it$.
Discarding this symmetry, one has larger freedom in formulating the
Lifshitz invariant equations of motion.

By analogy with our consideration above, in
order to construct a perfect fluid equations for an
arbitrary $z$ it suffices to introduce the conformal compensator $\rho(t)$. It
seems natural to keep the continuity equation intact, while the
Euler equation and the equation of state are modified as follows
\begin{eqnarray}\label{FLeq1}
\frac{\partial\varrho}{\partial t}+ \frac{\partial
(\varrho\upsilon_i)}{\partial x_i}=0,\quad {\cal
D}(\rho^\alpha\upsilon_i)=-\frac{1}{\varrho}\frac{\partial
p}{\partial x_i},\quad p=\nu\varrho^{1+\frac{z(4-\alpha)-2}{d}},
\end{eqnarray}
where $\alpha$ is an arbitrary constant. The compensator $\rho(t)$
transforms under the dilatations as follows
$\rho'(t')=e^{\frac{z\lambda}{2}}\rho(t)$. Note that equations (\ref{FLeq1}) are Lifshitz
invariant for arbitrary $\alpha$ and (\ref{FLeq}) is recovered for
$\alpha=0$. Because $\rho(t)$ cannot be removed by a field redefinition, it would be interesting to study its thermodynamic interpretation. We leave this issue for future study.

It is straightforward to construct the Hamiltonian which gives rise to the
fluid equations (\ref{FLeq1}) and simultaneously determines the
dynamics of the conformal mode $\rho(t)$
\begin{eqnarray}\label{HamPF}
H=\frac12\pi\pi+\int
dx\left[\frac{\varrho}{2\rho^{\alpha}}\tilde\upsilon_i\tilde\upsilon_i+V(p)\right].
\end{eqnarray}
The corresponding Poisson brackets read
\begin{eqnarray*}
\{\rho,\pi\}=1
\end{eqnarray*}
and
\begin{eqnarray}\label{PBPF}
\{\varrho(x),\tilde\upsilon_i(y)\}&=&- \frac{\partial}{\partial
x_i}\delta(x-y),\nn
\\
\{\tilde\upsilon_i(x),\tilde\upsilon_j(y)\}&=&\frac{1}{\varrho}\left(\frac{\partial\tilde\upsilon_j}{\partial
x_i} -\frac{\partial\tilde\upsilon_i}{\partial
x_j}\right)\delta(x-y).
\end{eqnarray}
Above we redefined the velocity vector field
${\tilde\upsilon_i}=\rho^\alpha\upsilon_i$, in terms of which the
Poisson brackets (\ref{PBPF}) coincide with those in \cite{JNPP}.

Because under the dilatation the Hamiltonian must scale as
$H'=e^{-z\lambda}H$, the value of the parameter $\alpha$ is fixed to
be $\alpha=\frac{2(z-1)}{z}$. Note that (\ref{HamPF}) has the same
structure as (\ref{MBH}) for the case of an infinite number of
particles.
The Hamiltonian (\ref{HamPF}) determines the dynamics of the
conformal mode
\begin{eqnarray}\label{ConfEqPF}
\dot\rho=\{\rho,H\}=\pi,\quad
\dot\pi=\{\pi,H\}=\frac{z-1}{z}\rho^{\frac{-3z+2}{z}}\int
dx{\varrho}\tilde\upsilon_i\tilde\upsilon_i
\end{eqnarray}
and reproduces the fluid equations of motion (\ref{FLeq1})
\begin{eqnarray}\label{LifEqPF}
\dot\varrho=\{\varrho,H\}=-\frac{\partial
(\varrho\upsilon_i)}{\partial x_i},\quad
\dot{\tilde\upsilon}_i=\{{\tilde\upsilon}_i,H\}=-\frac{\tilde\upsilon_j}{\rho^{\alpha}}\frac{\partial
\tilde\upsilon_i}{\partial x_j}-\frac{1}{\varrho}\frac{\partial
p}{\partial x_i},
\end{eqnarray}
provided the potential $V$ obeys the equation
\begin{eqnarray}\label{PotFeq}
\frac{1}{\varrho}\frac{\partial p}{\partial x_i}=\frac{\partial
V'_\varrho}{\partial x_i}.
\end{eqnarray}
Taking into account $p=\nu\varrho^{1+\frac{2z}{d}}$, one finds
$V=\frac{d}{2z}p$. Conserved charges corresponding to the dilatation, spatial
translation and rotation read
\begin{eqnarray*}
D=-\frac12z\rho\pi-\frac{1}{2}\int dx\varrho x_i\tilde\upsilon_i+zHt
,\quad C_i=\int dx \varrho \tilde\upsilon_i,\quad M_{ij}=\int dx
\varrho (x_i\tilde{\upsilon}_j-x_j\tilde{\upsilon}_i).
\end{eqnarray*}
They form the Lifshitz algebra under the Poisson bracket
\begin{eqnarray*}
\{H,D\}=zH,\quad \{D,C_i\}=-\frac12C_i,\quad
\{M_{ij},C_k\}=-\delta_{ki}C_{j}+\delta_{kj}C_{i}.
\end{eqnarray*}

Given the conformal mode $\rho(t)$, one can
construct a more general equation of state which will preserve the
Lifshitz symmetry
$$
p=\nu\rho^\mu\varrho^{1+\frac{z(2+\mu)}{d}},
$$
where $\mu$ is an arbitrary constant. This equation of state via (\ref{PotFeq})
determines the potential
$V=\frac{d}{z(2+\mu)}p$ and for $\mu=0$ it reproduces the third equation in (\ref{FLeq1}).

\section{Dynamical realizations of the Lifshitz group in field theory}\label{Sec7}

In this section, we focus on dynamical realizations of the Lifshitz group in field theory.
Let us consider again the
non-extended Lifshitz algebra
\begin{eqnarray*}
{[H,D]}&=&izH,\quad {[D,C_i]}=-\frac{i}{2}C_i
\end{eqnarray*}
and introduce the coset space element
\begin{eqnarray*}
g=e^{itH}e^{i{x}_iC_i}e^{iu(t,x)D}\times SO(d).
\end{eqnarray*}
The Maurer-Cartan one-forms
(\ref{formsd})
\begin{eqnarray*}
\omega_H=e^{-uz}dt,\quad \omega_D=du,\quad
\omega_i=e^{-\frac{u}{2}}d{x}_i
\end{eqnarray*}
hold invariant under the finite transformations
\begin{align*}
& t'=t+\beta, && u'(t',x')=u(t',x'), && {{x}'}{}_i={x}_i,
\\
& t'=e^{z\lambda}t, && u'(t',x')=u(t,x)+\lambda, &&
{{x}'}{}_i=e^{\frac{\lambda}{2}}{x}_i,
\\
& t'=t, && u'(t',x')=u(t,x), && {{x}'}{}_i={x}_i+a_i.
\end{align*}
Then one builds the invariant temporal and spatial derivatives as well as field combinations
\begin{align*}
& {\cal D}_t=\frac{d}{\omega_H}=e^{zu}\frac{d}{dt}, && {\cal
D}_i=\frac{d}{\omega_i}=e^{\frac12u}\frac{d}{d{x}_i},
\\
& {\cal D}_tu=\frac{\omega_D}{\omega_H}=e^{zu}\frac{du}{dt}, &&
{\cal D}_iu=\frac{\omega_D}{\omega_i}=e^{\frac12u}\frac{du}{d{x}_i}.
\end{align*}
Aftewards, one considers the most general equation, which is
quadratic in the derivatives
\begin{eqnarray}\label{ueqfl}
({\cal D}_t+a{\cal D}_tu){\cal D}_tu+\alpha({\cal D}_i+b{\cal
D}_iu){\cal D}_iu=2\gamma^2,
\end{eqnarray}
where $a,b,\alpha,\gamma$ are constants.

Introducing a conformal mode $\rho(t,x)$, one can build the invariant action
$$\frac12\int
dtdx\partial_t\rho\partial_t\rho,
$$
where we denoted $dx=dx_1...dx_d$. This corresponds to choosing
$\rho=e^{\frac{(2z-d)u}{4}}$ and fixes the parameters
$$
a=b=-\frac{2z+d}{4}.
$$
Then the equation (\ref{ueqfl}) takes the form
\begin{eqnarray*}
{\partial_t\partial_t\rho}+\alpha\rho^{-\frac{4(2z-1)}{2z-d}}[\partial_i\partial_i\rho-\frac{2(2z-1)}{2z-d}\rho^{-1}\partial_i\rho\partial_i\rho]=
\frac{(2z-d)}{2}\gamma^2\rho^{-\frac{6z+d}{2z-d}},
\end{eqnarray*}
and the corresponding action functional reads
\begin{eqnarray}\label{act}
S&=&\frac12\int
dtdx\left(\partial_t\rho\partial_t\rho+\alpha\rho^{-\frac{4(2z-1)}{2z-d}}\partial_i\rho\partial_i\rho
-\hat\gamma^2\rho^{-\frac{2(2z+d)}{2z-d}}\right),
\end{eqnarray}
where we redefined the coupling constant
$\hat\gamma^2=\frac{(2z-d)^2}{2(2z+d)}\gamma^2$.

Using Noether's
theorem one can construct the integrals of motion
\begin{eqnarray*}
H&=&\frac12\int
dx\left(\partial_t\rho\partial_t\rho-\alpha\rho^{-\frac{4(2z-1)}{2z-d}}\partial_i\rho\partial_i\rho
+\hat\gamma^2\rho^{-\frac{2(2z+d)}{2z-d}}\right),
\\
C_i&=&\int dx\partial_t\rho\partial_i\rho,
\\
D&=&\frac{2z-d}{4}\int dx\rho\partial_t\rho-\frac{1}{2}\int
d^dxx_i\partial_t\rho\partial_i\rho-zHt.
\end{eqnarray*}
When verifying conservation of over time, the identity
\begin{eqnarray*}
\partial_t(\partial_t\rho\partial_i\rho)&=&+\frac12\partial_i(\partial_t\rho\partial_t\rho)
-\frac{(2z-d)^2}{4(2z+d)}\gamma^2\partial_i\left(\rho^{-\frac{2(2z+d)}{(2z-d)}}\right)
\\
&&-\alpha\frac{(2z-d)^2}{(2z+d-2)^2}\partial_j\left(\partial_j(\rho^{-\frac{2z+d-2}{2z-d}})\partial_i(\rho^{-\frac{2z+d-2}{2z-d}})\right)\\
&&+
\frac12\alpha\frac{(2z-d)^2}{(2z+d-2)^2}\partial_i\left(\partial_j(\rho^{-\frac{2z+d-2}{2z-d}})\partial_j(\rho^{-\frac{2z+d-2}{2z-d}})\right)
\end{eqnarray*}
proves useful,
which holds provided the equations of motion are satisfied.

A scalar field theory described by the action functional (\ref{act})
corresponds to a particular case studied in \cite{IRS09}. For
$z=\frac12$ and $\alpha=-1$, the action (\ref{act}) describes the
relativistic conformal scalar field.

\section{Conclusion}\label{S8}

To summarize, in this work we constructed new dynamical realizations
of the Lifshitz group in mechanics, hydrodynamics, and field theory.
The main ingredient of our consideration was the conformal compensator $\rho(t)$.
Using the group-theoretic approach, in which $\rho(t)$ naturally arises as
a partner of the dilatation, the following new results were
obtained. It was shown that one-dimensional mechanics with the Lifshitz symmetry
by a proper field redefinition can be linked to $so(2,1)$ conformal mechanics for arbitrary $z$, not
just for $z=1$ as in \cite{AG22}. In multidimensional mechanics, a
complete Lagrangian formulation was attained, in which $\rho(t)$ and $x_i(t)$ are treated on equal footing.
Many-body examples were also constructed and studied.
Generalized Calogero-like models involving arbitrary $z$ were built.
Two higher derivative examples, in which the Lifshitz algebra is extended by the Galilean boost
generator as well as constant accelerations,
were formulated. Perfect fluid equations with the Lifshitz
symmetry and its Hamiltonian formulation were built. It was demonstrated that the
group-theoretical approach can also be used in order to construct
Lifshitz-type scalar field theories.

As a possible further development it would be interesting to study
in more detail the geometric aspects of the extended Lifshitz
algebra(\ref{EzLal}) as well as to construct its dynamical
realizations in hydrodynamics and field theory. Causality of the
field theory (\ref{act}) deserves a separate investigation.

\section*{Acknowledgements}
This work was supported by
the Russian Science Foundation, grant No 23-11-00002.

\end{document}